# NETWORK-CENTRIC OPTIMAL HYBRID SENSING HOLE RECOVERY AND SELF-HEALING IN IPv6 WSNs

Kwadwo Asante [1], Yaw Marfo Missah [2], Frimpong Twum [2] and Michael Asante [2]

[1] Department of Information Technology Education, AA-MUSTED, Kumasi
[2] Department of Computer Science, KNUST, Kumasi, Ghana.

## ABSTRACT

*In our earlier work, Network-Centric Optimal Hybrid Mobility for IPv6 wireless sensor networks, in which the work sought to control mobility of sensor nodes from an external network was proposed. It was a major improvement on earlier works such as Cluster Sensor Proxy Mobile IPv6 (CSPMIPv6) and Network of Proxies (NoP). In this work, the Network-Centric optimal hybrid mobility scenario was used to detect and fill sensing holes occurring as a result damaged or energy depleted sensing nodes. Various sensor networks self-healing and recovery, and deployment algorithms such as Enhanced Virtual Forces Algorithm with Boundary Forces (EVFA-B); Coverage - Aware Sensor Automation protocol (CASA); Sensor Self-Organizing Algorithm (SSOA); VorLag and the use of the use of anchor and relay nodes were reviewed. With node density thresholds set for various scenarios, the recovery efficiency using various parameters were measured. Comparably, our method provides the most efficient node relocation and self-healing mechanism for sensor networks. Compared to Sensor Self-Organizing Algorithm (SSOA), Hybrid Mobile IP showed superiority in coverage, shorter period of recovery, less computational cost and lower energy depletion. With processing and mobility costs shifted to the external network, Hybrid Mobile IP extends the life span of the network.*

## 1. INTRODUCTION

The aim of sensor networks deployed purposely for environmental monitoring is to provide full and uninterrupted coverage of the sensing field. To improve the service availability, the sensors must automatic recover sensing holes resulting from damages and other service interruptions [1]. Network recovery mechanisms should be able to discover, diagnose, and react to network disruptions. Self-healing components can detect system failures such as energy depletion and blackout due to sensor node damages and activate corrective actions based on defined policies to recover the network [18]. To achieve this, the network must employ efficient mobility and node relocation mechanisms to recover from failures. The ability of sensor nodes to recover from sensing holes has been a subject of intense research over the last few years. The major hurdle has been energy depletion of sensor nodes due to mobility and excessive computations which reduces the life span of the nodes and the network at large [11]. Various recovery methods have been proposed but these methods have not been efficient largely because of cost of computation by sensor nodes and their ability to withstand the mobility related energy depletion. To overcome this problem, our earlier work [30] proposed Network-Centric Optimal Hybrid Mobility Scheme for IPv6 Wireless Sensor Networks, in which mobility of mobile nodes and all associated computation were controlled from the network side. This enables sensor nodes to retain much of their energy for sensing and node registration while mobility and other related computations are handled by the network. Compared to existing healing and recovery mechanisms, such as Sensor Self-Organizing Algorithm (SSOA), Hybrid Mobile IP showed superiority in coverage, detection of the existence of sensing holes, shorter period of recovery, less computational cost and lower energy depletion.

## 2. RELATED WORKS

There have been a lot of work with regards to sensor network deployment and network management, topology control and recovery of sensor networks. Our review extends from network recovery mechanisms to mobility related issues.

### 2.1. VorLag ( Voronoi-Laguerre diagrams)

Each sensor uses the sensor location over the AoI and their related sensing radii to construct VoronoiLaguerre Diagrams. The information gathered from this diagram is used to determine its future movements and the presence of coverage holes [11]. Each sensor broadcast their location and with respect to its neighbors, calculates the radial axes it generates based on the available information on neighboring sensors and use it to construct its Voronoi polygon. The algorithm iterates and after each round, sensors construct their polygon and evaluates the existence of coverage holes within their polygons and movements decisions are made accordingly until node density is fairly distributed over the AoI . No polygon is generated by a sensor in an overcrowded area; thus, such sensor does not move. The Vorlag algorithm determines the target location of a sensor within its Laguerre polygon whose Euclidean distance from the farthest vertex is minimize [8].

### 2.2. Coverage - Aware Sensor Automation protocol (CASA)

Defined by [3], Coverage-aware optimal sensors deployment is a process of achieving the coverage requirement of an application by determining the optimal locations of sensors in a network field. The ability of a WSN to detect and recover sensing holes is a sub-problem of deployment protocols. [2] suggests the use of mobile sensors to fill sensing holes by adapting their positions on the sensing field. In order to minimize the number of sensors required to cover sensing holes [5] proposed the Maximum Coverage Sensor Deployment Problem ((MCSDP) which is modelled as a constrained optimization problem and aims at finding the minimum number of sensors required to achieve maximum coverage of the sensing field. CASA employs two set of protocols; Enhanced Virtual Forces Algorithm with Boundary Forces (EVFA-B) and Sensor Self-Organizing Algorithm (SSOA)

### 2.3. Enhanced Virtual Forces Algorithm with Boundary Forces (EVFA-B)

EVFA-B is inspired by combined idea of potential field forces and disk packing theory [10]. Two neighboring sensors behave like two electrostatic charged particles, exerting force of attraction or repulsion on each other based on a predefined distance between them [9]. A distance threshold is set by the sensing range of two sensor nodes and incorporates a boundary force which denotes the virtual force acting from the monitored boundaries, which reduces sensing outside the sensing field. EVFA-B is a global deployment algorithm and all computation are done by the sensor nodes. This leads to excessive depletion of node energy and costly mobility overheads, causing more node failures and shortening network lifespan. Thus, EVFA-B algorithm is used seldomly for global re-deployment to cover sensing holes [14].

## 2.4. Sensor Self-Organizing Algorithm (SSOA)

The **Sensor Self-Organizing Algorithm (SSOA)** is an algorithm designed to optimize the deployment, coordination, and management of sensor networks in a **self-organizing** manner. It aims at ensuring that sensors in a network can organized autonomously and adjust their configurations to recover sensing holes without human intervention. SSOA seeks to repair sensing holes by relocating sensors around the sensing holes. Proposed by [6], it involves the selection of repairing sensors and mobility of these sensors. According to [4], SSOA is limited to a single-tier movement due to complex computations with multiple-tier movements of recovering sensors. However, full recovery requires more tier movements of recovering sensors. According to [12], SSOA provides an effective self-healing capabilities where faulty sensors are generally distributed across the network. In situations where sensor faults are concentrated at a particular location resulting in reduced sensor area, the re-deployment algorithms such as EVFA-B may be required.

## 2.5. The use of Anchor and Relay Nodes

**Local Selection of Rescue Sensors-** The first attempt to carry out sensor hole restoration is the use of neighboring sensors. To obtain the most desirable coverage improvement, the choice of rescue sensors and movement strategy must be adopted to maintain communication with other sensors, thus, a distance threshold with reference to the communication range of the sensors must be defined. According to [13], the use of nonadjacent nodes, with certain level of overlapping degrees provides most efficient recue coverage.

[7] proposed **Ping-Pong Free Advanced and Energy Efficient Sensor Relocation for IoTSensory Network** in which they proposed relocation protocol which showed a uniform and energy-efficient sensor movement within a cluster zone upon request. In this proposal, sensor nodes are grouped in zonal clusters with cluster heads and relay nodes serving as edge routers as shown in fig1 below. When the number of sensing nodes are insufficient in a cluster, sensing hole occurs and the sensing hole header request for more sensor nodes from the nearby cluster zones through the relay node. The relay node uses the relocation protocol [15].

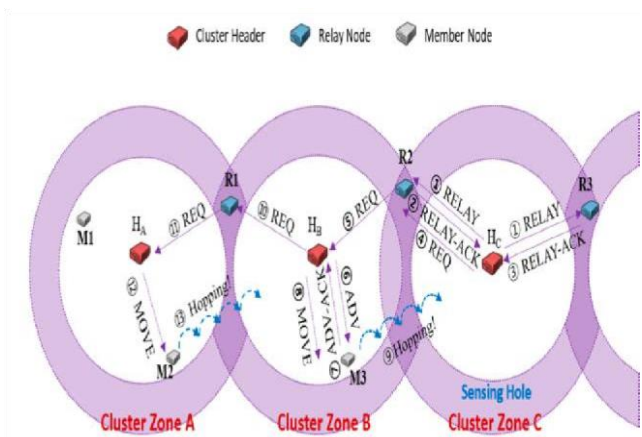

Fig.1. Hopping sensors relocation [7]

**Smart Node Relocation (SNR) -** SNR uses SYN messages broadcast periodically from neighbouring nodes to detect node failure. When a node fails to respond a SYN message from neighbouring nodes, it is considered dead. Neighbouring nodes triggers mobility related signalling and moves towards the direction of the holes created to cover up [8]. It focuses on moving minimal number of nodes in the process of restoring connectivity to the network using heartbeat messages generated by the neighbouring nodes [18].

## 2.6. Network Related Mobility Techniques

All the above relocation and self-healing techniques employs node mobility to recover sensing holes and are prone to severe energy loses due to the mobility and related signaling. This work therefore took a summary review of network assisted mobility techniques. .

### 2.6.1. Sensor Proxy Mobile IPv6

The major bottleneck in the SPMIPv6 [22] was the LMA being the central point of failure. This is as a result of the LMA keeping the binding cache entries and being the main route for all messages and data packets. Thus, the LMA was placed in a heavy-duty situation and other parameters such as route optimization and latency were heavily affected. A lot of non-optimized routes were created since exchange of information between nodes in the same domain still passed through the LMA. The SPMIPv6 was hardware intensive and expensive to deploy [27]. The setup is depicted in fig 2. below.

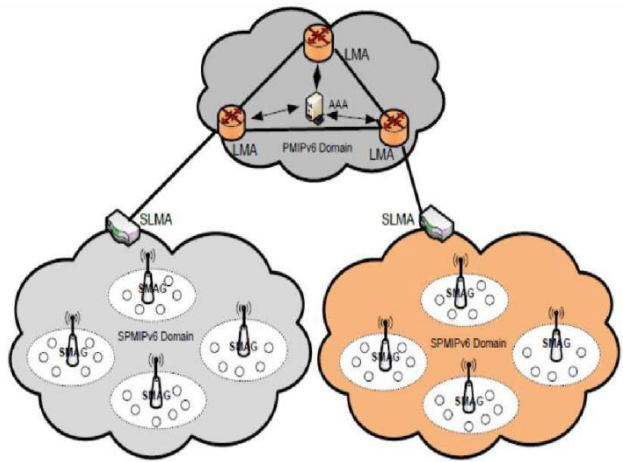

Fig. 2. SPMIPv6

### 2.6.2. Cluster Sensor Proxy Mobile IPv6

The Cluster Sensor Mobile IPv6 [31] showed much improvement over the SPMIPv6. In addition to the MAG and LMA, the CSMIPv6 introduced the Head MAG to handle mobility management within each cluster, hence reduced the over dependency on the LMA. Again the head MAG integrated AAA and provided route optimization for both inter and intra cluster communication. The head mag reduced the signaling cost greatly by providing enough buffering for Mobile Nodes during the hand-off process [29]. However, the CSMIPv6 employs a static trees-based backbone structure, which is very complex and prone to the same major drawback in the SPMIPv6, single point of failure because, it still uses the LMA as single anchor to the external network.

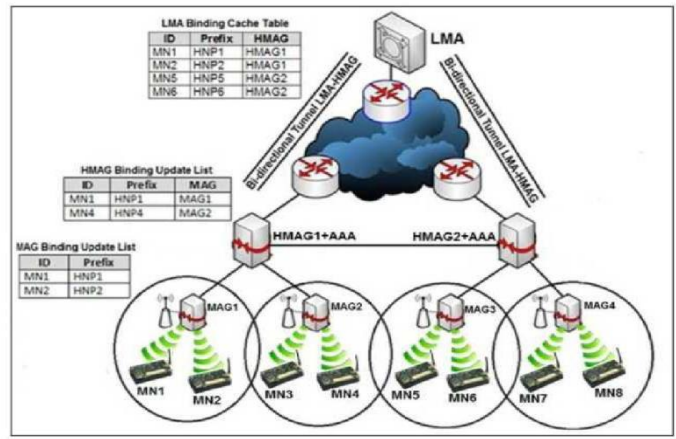

Fig.3. CSMIPv6

### 2.6.3. Network of Proxies

Design for critical scenarios, Network of Proxies (NoP) was constructed as a wireless mesh network overlay constituted by resource unconstraint nodes which handle on behalf of the sensor nodes all the high energy consuming and time constraint processes [25]. NoP does not have the bottlenecks of the SPMIPv6 and CSPMIPv6 and improved the hand-off process considerably. However, the process takes place between the sensor nodes and the proxy network. Mobility is applied to only the sink nodes while the sensor nodes remain static [21]. There is therefore a possibility of sensing hole occurring. The architecture is complex and expensive and it is implemented for only critical /special scenarios

## 3. THE PROPOSED SOLUTION

### 3.1. Proposed Setup

OMNeT++ simulator was used to investigate node relocation and self-healing of a mobile sensor network using a hybrid network. Our setup consisted of two separate networks: the Wireless sensor network and an enterprise network / internet backbone which comprise of ethernet and wireless links. The two setups were linked by a middleware which enabled the network to perform node mobility, address assignment and handoff processes on behalf of the mobile node. The WSN was divided into zones and each zone was further divided into clusters. With this simplified architecture, our solution sought to eliminate the hardware complexities in the CSPMIPv6. Our setup is shown in fig.4 below.

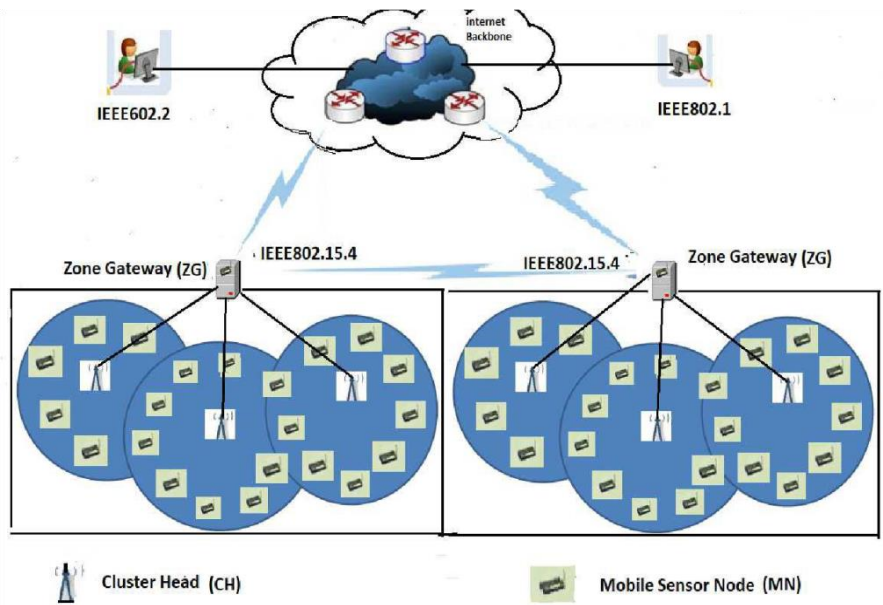

Fig. 4: Proposed Hybrid Mobile IPv6

**IP-Sensor Nodes, SN** – The sensor node was embedded with the tiny TCP/IP as well as IPv6 protocol stacks and IEEE 802.15.4 interface with adaptation layer. It has an environmental sensing capability and able to implement end-to-end communication. It forwards sensed information to the cluster head (CH) either directly or through a single hop with neighboring node. Each node is embedded with address, home network prefix and a flag bit indicating sensor proxy registration.

**The Cluster Head, CH** – The CH incorporated the 6LowPAN adaptation layer which performed the task of transmitting IPv6 packets over IEEE 802.11 and IEEE802.3 links. It acted as a sink node and provided an access gateway routing for sensor nodes. CH activates mobility -related signaling with the zonal gateway (Gz) as well as act as an anchor node to control the boundaries of the mobile sensor nodes. In addition, it performs neighbor discovery and multiple address detection. RPL was used as the default routing protocol.

**The Zone Gateway (Gz) -** Its job is to sustain accessibility to the IP sensor nodes while the nodes move within or outside the zone by acting as the main gateway to the external network. It interfaces the Sensor network and the external network and contains all the information embedded in each cluster head within a zone. Provided with sufficient memory and power supply for processing, it is also configured to provide authentication and secure mobility for sensor nodes. Gz is a gateway router and also acts as a DHCPv6 relay agent to the sensor networks and also configured to run RPL.  Fig 3.2 below shows the operational architecture of the proposed system

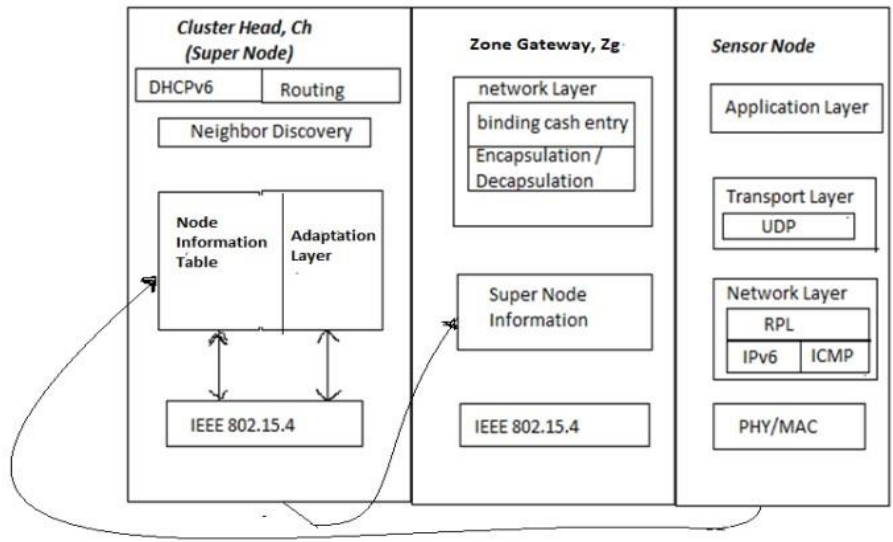

Fig. 5: The operational architecture of the proposed system

## 3.2. Communication in the Proposed System

When a sensor node arrives in a cluster from another cluster, it undergoes a registration process as follows:

Step 1: CH detects a new mobile node when it arrives in a cluster
Step 2: CH request the identity of the new node by sending ID-REQ to the Zonal Gateway Gz.
Step 3: Gz performs fast authentication on the new node (if the node is from a cluster within the same zone).
Step 4: Gz forwards request acknowledgment (REQ-ACK) to the CH. The REQ-ACK is embedded with the node information including its position.
Step 5: CH registers the new node
Step 6: CH updates its cache entries with the new node information (including its new position)
Step 7: Gz updates its cache entries with the new CH information
Else, if node has a different home prefix (from a different cluster),
Go to step 2
Step 3: Gz sends an ID-REQ to neighboring Gz, requesting the identity of the new node (using its home prefix)
Step 4: The neighboring Gz replies with REQ-ACK containing the node information
Step 5: The REQ-ACK is forwarded to the CH that made the initial request
Step 6: The CH registers the new node using its new home prefix and computes its coordinates
Step 7: CH updates its cache entries with the new node information
Step 8: Corresponding Gz updates its cache entries with the new node information.

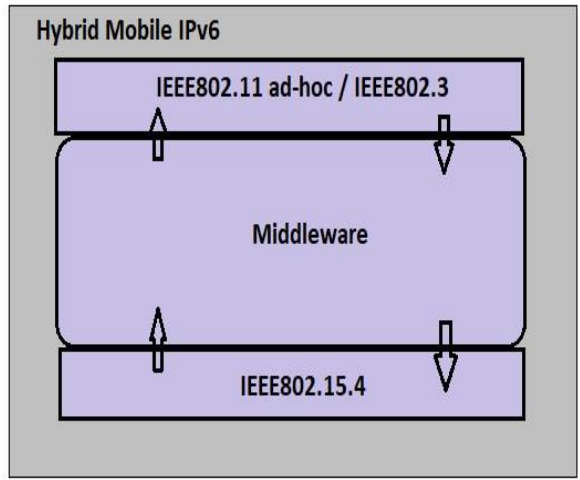

Fig.6: Interfaces of the proposed system

### 3.3. Setup Parameters

Initially, the nodes were deployed using the Virtual Force Method with a boundary force implemented by the CHs. The sensor nodes were spread evenly and based on their distances and attraction from the CHs, they were grouped into clusters. The clusters were grouped into zones by allowing equal circles to be paired in squares. Table 1 below gives the parameters of our setup.

Table 1: Setup parameters

| PARAMETER | VALUE |
| --- | --- |
| Network area | 220 x 220m |
| No of sensing holes | variable |
| Packet size | 2KB |
| Initial energy | 6J , 30J |
| Transmission power | $1.18 \times 10^{-3}$W |
| Transmission range | 5m |
| Node densities per zone | variable |
| Physical layer | IEEE802.15.4 |
| Network layer | RPL |
| Transport layer | UDP |

To identify coverage holes, a grid-based approach was used, where the monitored area was divided into grids and each grid cell was analyzed to determine if it is adequately covered by sensor nodes. The middleware used the following algorithm to detect sensing holes and initiate remedial action.

**Algorithm for identifying coverage holes.**

Inputs:
- `nodes`: A list of sensor nodes, each with coordinates.
- `grid`: A 2D grid representing the sensing area.

Outputs:
- A list of coverage holes.

Steps:

1. Initialize Coverage Map:
   - Create a 2D list called `coverageMap` with the same dimensions as `grid`.
   - For each cell in `coverageMap`, set the value to `false` (indicating that it is initially uncovered).

2. Mark Covered Cells:
   - For each `node` in `nodes`:
     - Retrieve the coordinates `(x, y)` of the node.
     - Set `coverageMap[x][y]` to `true` (indicating this cell is covered by the node).

3. Identify Coverage Holes:
   - Create an empty list called `coverageHoles`.
   - For each `cell` in `coverageMap`:
     - If the cell is `false` (not covered):
       - Add the cell to `coverageHoles` (mark this cell as a coverage hole).

4. Return Result:
   - Return the list `coverageHoles` containing all identified coverage holes.

Auxiliary Function:

Function: Initialize Grid
- Input: `grid`
- Output: `coverageMap`
- Steps:
  1. Create an empty list `coverageMap`.
  2. For each `row` in `grid`:
     - Create a new empty list `newRow`.
     - For each `cell` in `row`:
       - Append `false` to `newRow`.
     - Append `newRow` to `coverageMap`.
  3. Return `coverageMap`.

Function: Mark Coverage Hole
- Input: `cell`
- Purpose: Implement any necessary logic to handle or log the identified coverage hole.

| **Sensing hole recovery algorithm** |
|---|
| Algorithm: Cluster Management<br><br>Inputs:<br>- A list of \`clusters\`, each containing a list of \`sensors\`.<br>- A \`Max_threshold\` for the maximum number of sensors allowed in a cluster.<br>- A 'Min_threshold` for the minimum number of sensors allowed in a cluster.<br><br>Data Structures:<br>- Cluster: Contains a list of sensors.<br>- Sensor: Represents a sensor (details not specified).<br>Steps:<br>1. Initialize Cluster Manager:<br>  - Create a \`ClusterManager\` with a specified \`Min_threshold\`, Max_Threshold.<br>  - Initialize an empty list of \`clusters\`.<br>2. Add Sensor to Cluster:<br>  - Define a function to add a \`sensor\` to a specified \`cluster\`.<br>  - Append the sensor to the \`sensors\` list of the selected cluster.<br>3. Remove Sensor from Cluster:<br>  - Define a function to remove a \`sensor\` from a specified \`cluster\` by index.<br>  - Remove the sensor from the \`sensors\` list at the specified index.<br>4. Move Sensor Between Clusters:<br>  - Define a function to move a \`sensor\` from one cluster to another.<br>  - Retrieve the sensor from the source cluster.<br>  - Add the sensor to the target cluster.<br>  - Remove the sensor from the source cluster.<br>5. Manage Cluster Load:<br>  - For each cluster:<br>    - If the number of sensors exceeds the \`Max_threshold\`:<br>      - Find a target cluster with the least number of sensors (see step 6).<br>      - Move a sensor (e.g., the first one) to the target cluster.<br>    - If the number of sensors is below the \`Min_threshold\`:<br>      - Find a source cluster with the most sensors (see step 7).<br>      - Move a sensor from the source cluster to the current cluster if a suitable source exists.<br>6. Find Target Cluster:<br>  - Initialize \`minSize\` to a large value.<br>  - Iterate through the clusters:<br>    - If the current cluster has fewer sensors than \`minSize\`, update \`minSize\` and record the index as the target cluster.<br>  - Return the index of the target cluster.<br>7. Find Source Cluster:<br>  - Initialize \`maxSize\` to zero.<br>  - Iterate through the clusters:<br>    - If the current cluster has more sensors than \`maxSize\`, update \`maxSize\` and record the index as the source cluster.<br>  - Return the index of the source cluster.<br><br>8. Main Execution:<br>  - Create an instance of \`ClusterManager\` with the desired thresholds.<br>  - Add clusters to the manager.<br>  - Add sensors to the appropriate clusters.<br>  - Periodically call the \`manageClusterLoad\` function to ensure balanced load among clusters. |

After discovering sensing holes, in a particular cluster, each sensor in the cluster adjusts its position using the vector, V, given by

$$V = \left[\begin{pmatrix} vx \\ vy \end{pmatrix}\right]$$

The initial and the new position of sensor nodes is now given by

$$\left(\begin{bmatrix} xf \\ yf \end{bmatrix}\right) = \begin{bmatrix} xfi \\ yfi \end{bmatrix} + \begin{bmatrix} vx.\ \Delta t \\ vy.\ \Delta t \end{bmatrix}$$

Rf =

Vx represents the rate of change of in the x-direction

- Vy represent the rate of change in the y-direction This means:
- The new *x*-position of the sensor $x_f$ is given by $x_f = x_{fi} + vx.\ \Delta t$
- The new *y*-position of the sensor $y_f$ is given by $y_f = y_{fi} + vy.\Delta t$ Updating sensor position:
- *Initial Position* $Rf_i$: the position of the sensor at the start (before the adjustment), the Rfi is represented as ($xf_i$, $yf_i$) in the cartesian plane.
- *Velocity vector* **V**: the velocity vector V determines how much the position changes over each time step. It is a two-component vector, where:

*The time step, Δt*, represents the time taken for the sensor node to move from its initial position to the final position.

The *final position, $R_f = R_i + V.\ \Delta t$*

The above equations represent the rate of change of position of the sensor node. It specifies both the speed and direction of the node's movement. The middleware implements a **ClusterManager Class** which manages clusters and implements methods to add, remove, and move sensors between clusters based on the defined thresholds.

**The Threshold Management** function iterates through clusters and checks if they exceed Max_Threshold, or fall below the Min_threshold. If a cluster exceeds the Max_threshold, it moves sensors to another cluster. If it falls below the Min_threshold, it retrieves sensors from another cluster. Below is the algorithm to simulate sensor movement and sensing hole recovery process.

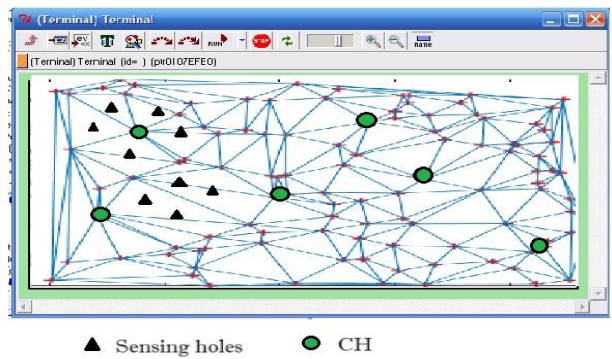

Figure 7: Cluster deployment

### 3.4. Simulation Results

The Hybrid Mobile IP algorithm was run at the network side to detect sensing holes, initiate recovery and to controlled node mobility. It was run alongside the SSOA to compare the effectiveness of both protocols.

*Recovery Period, $T_r$* – A number of sensing holes were created with the intention of investigating the recovery period and the factors that have an impact. In other words, how fast can a cluster recover its blind spots. With a nodes population of 674 per zone and five clusters in each zone, with a cluster Min_threshold, $\Phi_{mim}$ = 124 and Max_threshold, $\Phi_{max}$ = 135, and sensor energy was set to its default of 6J. The number of sensing holes and their corresponding recovery time, Tr were measured. These sensing holes were uniformly distributed in a single cluster. The variation was as shown in the graph in fig.8 below.

Again, the *Recovery Period, $T_r$* for recovering 20 sensing holes was measured by varying the node density per cluster. The number of sensing holes was defined to be the number of sensors nodes required to achieve the minimum threshold in the test cluster. The node density of the neighboring clusters was initially deployed to be above the maximum threshold, causing nodes to migrate to the test cluster. Fig 9. below shows the pattern.

*Average distance moved by recovery sensors* - With node mobility controlled by the external network, there was no need to set a threshold for distance moved by a particular node during relocation. However, in order to maintain the topology, each node bordering the sensing hole moves a small vector displacement $\Delta x$, towards the sensing hole repeatedly, until the sensing hole is recovered. Our setup was also used to investigate number of sensing holes against the average distance moved by nodes for both protocols. the outcome is as shown in fig.10 below.

*Percentage Coverage* – We measured the percentage coverage by initiating the minimum threshold of 600 and maximum threshold of 750 per zone (120 and 150 per cluster respectively). A number of sensors were allowed to populate the clusters as the percentage coverage for the test cluster was measured for both protocols. The outcome was as shown in fig.11 below.

*Computational Cost*– We also examined the computational cost for our set up based on Data size, Rate of processing. bandwidth and overhead costs.

$$C = k1.D + k2\frac{D}{R} + k3\frac{D}{B} + k4.O$$

k1 = cost per unit of data size
k2 = *cost per byte per operation per second*.
k3 = cost per unit of data per bandwidth   k4 = cost per unit of overhead
D = data size
R = Rate of processing
B = bandwidth
O = overhead cost

The curves in fig.12 was obtained for both protocols when the number of node density was varied and corresponding computation cost were calculated using the simulator.

**3.4.1. Energy Consumption Compared with SSOA**

Our setup was used to compare the energy depletion by nodes during network recovery. With initial energy of nodes set to 30J; cluster Min_threshold of 120 and Max_threshold of 150 per cluster, the number of sensing holes were varied against the average energy consumed by nodes in recovering sensing holes. Energy consumption for Hybrid Mobile IPv6 and SSOA compared (the maximum simulation time was 2mins and 30secs and it involved global redeployment of the nodes in the cluster) is shown in fig.13.

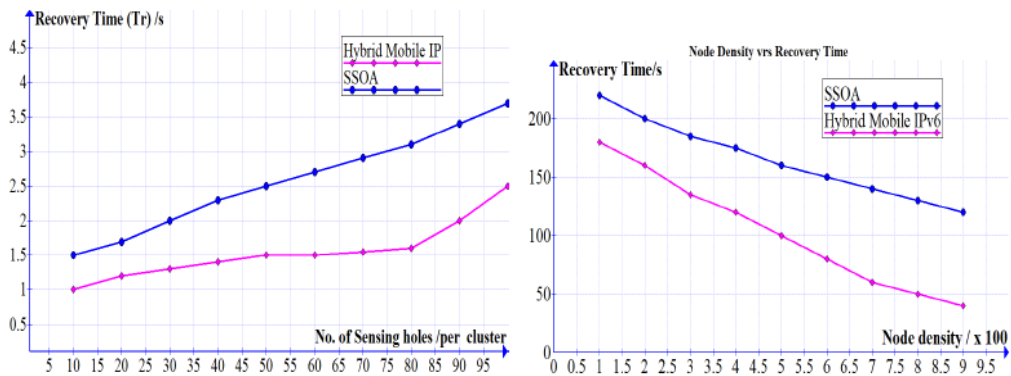

Fig. 8 : No. of sensing holes vrs Recovery time  Fig. 9: Node density against Recovery time

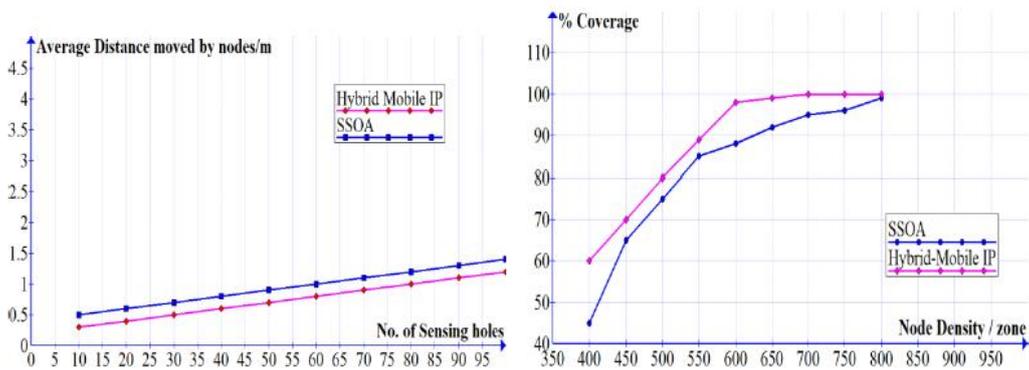

Fig.10 : No. of sensing holes vrs distance moved Fig.11; Node Density vrs % Coverage

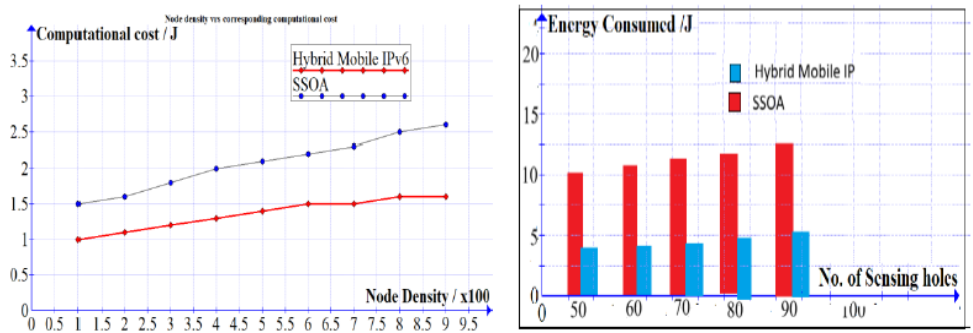

Fig.12 computational cost against node density     Fig.13 No. of sensing holes vrs energy consumed

## 4.1. Discussion of Simulation Results

As number of sensing holes increased, the recovery time increased steadily but slightly as shown in fig 8. However, it is observed that, the time taken to recover increased sharply when the number of sensing holes exceeded 80. This is because the number of remaining sensors reached its minimum threshold to provide coverage within a cluster. when reduced further, sensors from neighboring clusters migrate to provide coverage and they must go through authentication and registration processes, thus, the delay in recovering the sensing holes. From the experiment, the average time required to cover sensing holes was uniform until remaining nodes density exceeded the minimum threshold, and additional recovery nodes cross from neighboring clusters to recover sensing holes. At this point, positions of nodes in neighboring clusters only, are re-constructed. Compared to SSOA, where the entire network was reconstructed, Hybrid Mobile IP did 50% on the average of the recovery time of SSOA, making it more efficient.

In fig.9, $Tr$ declines steadily as node density increases. This is as a result of decreased node intervals as more sensor nodes are introduced. Nodes overlapping and few nodes are moved at short distances to cover up sensing holes. In SSOA, nodes under only a single tier during recovery. If the algorithm still detects a sensing hole, it deploys EVF-B to reconstruct the topology. On the other Hybrid Mobile IP deploys a looping mechanism by increasing the node position by a small distance at each loop, until the test cluster is fully recovered. As such Hybrid Mobile IP outperformed SSOA by an average of 40% in terms of how fast the sensing holes were recovered. . When node density decreased below the cluster threshold, the entire nodes in the cluster have to adjust their positions to cover up. Below the cluster threshold, nodes migrate from neighboring clusters, undergo through registration, thus the sharp delay in both protocols.

The average distance moved by sensor nodes to recover sensing holes increased steadily as the number of sensing holes increased. In effect, each node is allowed to move small distance, repeatedly until the sensing holes are recovered as depicted in fig 10. There was not much difference between both protocols with regards to the average distances moved to recover sensing hole. It was observed that, the average distance moved rather depends on the sensing range of the nodes, but not the protocols. At lower number of sensing holes, Hybrid Mobile IP performs better than SSOA but SSOA portrayed a steady increase as compared to Hybrid Mobile IP which showed a sharp increase in distance as nodes migrate from neighboring clusters.

In fig.11, we measured the percentage coverage as node density is increased steadily. In normal sense, coverage increases as more nodes are introduced. However, it is observed that Hybrid Mobile IP achieved much more coverage in the test cluster as compared to SSOA. This is as a result of Hybrid Mobile IP achieving more uniformity of distribution compared to SSOA. At node density 400 (100 below the threshold), SSOA achieved inly 45% while Mobile IP achieved 60% coverage. At the minimum threshold, Hybrid Mobile IP achieved almost 99% while SSOA achieved 88% covered. This is because, while Hybrid Mobile IP deploys uniformly in rapid succession, SSOA requires multiple tiers to fully cover the test field as exist in EVFAB. This result in time consuming and a lot of energy usage.

The amount of computations performed by the network in discovering sensing holes and recovering them is huge and can affect the efficiency and the life span of the network. In SSOA, all the computations are carried out by the nodes. in Hybrid Mobile IP, processing cost, overhead cost and mobility related cost are carried out by the network, thereby taking a lot of the stress from the sensor nodes. This allows the sensors to maintain their energy, thereby extending the lifespan of the network. Again, since the external network is able to contact the exact sensors required to move to cover sensing holes, few sensors are required to exchange packets at any point in time, except when large number of sensing holes occur and the network needs to be reconstructed. In fig.12, hybrid Mobile IP did performed an average of 50% better than SSOA.

In fig.13, the energy consumption by nodes during network recovery was investigated. It was observed that the average energy depletion was 13% compared to SSOA Protocol which gave an average of 38% for the same simulation period. This is because, in SSOA, the mobility overhead and other related computations such as node position were dependent on the node. In Hybrid Mobile IP, the mobility and other related computations are shifted to the external network. The node energy is used only in sensing the environment and performing authentication and other related signaling. Thus, our method reduced energy consumption by node and increased the life span of the network. Sensor movements were fairly controlled by the network during recovery; thus, energy depletion was very minimally spread across all sensors in a cluster.

So far, the Hybrid Mobile IPv6 employs the grid and threshold management approach to detect coverage holes. Both the threshold management and the cluster management classes enable clusters to maintain minimum and maximum number of sensor nodes, thereby preventing concentration of sensors in a particular cluster. Position Adjustment Based on Velocity Vectors was used to adjust node positions to recover sensingholes.

### 4.2. Conclusion

This work extends the capabilities of the Hybrid Mobile IPv6 to cover network self-healing and sensing hole recovery. Hybrid Mobile IPv6 has proved to be more efficient in detecting and recovering sensing holes by moving nodes using the vector approach towards void spaces in a sensor network. Hybrid Mobile IP is more efficient compared to the existing methods of sensor self-healing mechanisms. Our method presents the best of sensor nodes distribution mechanism that offers the most uniformity across the sensing fields since it operates within a minimum and maximum threshold. It initiates automatic movements of recovery without any human intervention once these thresholds are met. Compared to Sensor Self-Organizing Algorithm (SSOA), Hybrid Mobile IP showed superiority in coverage, shorter period of recovery, less computational cost and lower energy depletion. With processing and mobility costs shifted to the external network, Hybrid Mobile extends the life span of the network. Our work was implemented both in C++ and in python using the omnetpy and the results were the same.

### 4.3. Further Works

In this work, the underlying protocol was RPL. Further work is required with different LLNs protocols such as Ad-hoc On-Demand Distance Vector Routing and other variants of RPL, which we intend to explore in our next work. Again, optimal IPv6 addressing for large scale sensor networks must be explored to enable easy tracking of a particular node in rapidly changes WSN topologies. We used only one test cluster to test our parameters. Though nodes migrated successfully to recover sensing holes, the behavior of the network could not be predicted with sensing holes scattered across multiple clusters in large quantities. Again, we intend to employ AI tools such as Genetic Algorithms, Particle Swarm Optimization, or Reinforcement Learning to enhance optimisation and autonomous movements.

## AUTHORS

**Kwadwo Asante** is a lecturer at the Department of Information Technology Education at the Akenten AppiahMenkah University of Skills Training and Entrepreneurial Development (AA-MUSTED), Ghana. He has his specialty in computer networks, distributed system and IoT. He currently holds an M.Phil in Information Technology and A Ph.D. student at KNUST.

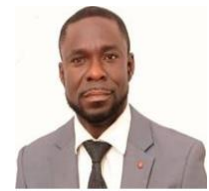

**Yaw Marfo Missah (Ph.D.)** is a Senior Lecturer at the Department of Computer Science, Kwame Nkrumah University of Science and Technology (KNUST ), Kumasi Ghana. He has researched in vast areas in computer science including areas in Integrating Expert Systems in Education, Neural Expert Systems in Education, Cryptographic Schemes in Cloud Data, Psychological and Social Interactions in AI.

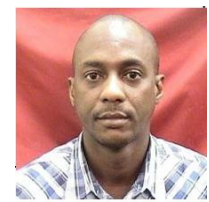

**Prof. Frimpong Twum (Ph.D.)** is an Associate Professor of Computer Science at the Department of Computer science, KNUST, Ghana. He has several years of experience in research and has published in wide areas of computer science including Computer Network Security, Machine Learning and AI Systems, Computer Vision and Image Processing, Cloud Computing, Software Engineering, Computer Programming and Algorithms

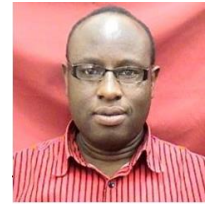

**Professor Michael Asante (Ph.D.)** is currently a professor in Computer Science at the department of Computer Science at the Kwame Nkrumah University of Science and technology, Kumasi, Ghana. He has researched and published widely in areas including Data Communication, Computer Security, Distributed Systems, Computer Networking, Infrastructure Security.

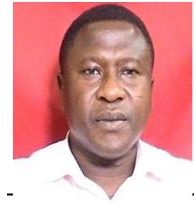